# Measurement of delayed fluorescence in $N_2^+$ with streak camera

Ali Rastegari and Jean-Claude Diels

*Center for High Technology Materials, University of New Mexico, Albuquerque,*
*New Mexico, NM 87106, USA*

Lee R. Liu

*Department of Physics, Harvard University, Cambridge, Massachusetts, 02138, USA*[a]

Ladan Arissian

*National Research Council of Canada, Ottawa, Canada*[b]

Using a streak camera, we directly measure time- and space-resolved dynamics of $N_2^+$ emission from a self-seeded filament. We observe characteristic signatures of superfluorescence even under ambient conditions and show that the timing of the emitted light varies along the length of the filament. These effects must be taken into consideration for accurate modelling of light filaments in air, and can be exploited to engineer the temporal profile of light emission in air lasing.

**Keywords:** Light filaments, Streak camera, Remote sensing, Spectroscopy, Air lasing

## I. INTRODUCTION

Creating a high brightness source in atmosphere is of interest for remote detection of gases and electromagnetic sensing. Light filaments are stabilized by the balance between self focusing and defocusing in air[1]. Their location can be controlled using temporal focusing[2]. Light filaments host enriched spectral contents such as supercontinuum generation and ion emission.

The transitions of the nitrogen cation $N_2^+$ are of particular interest for this application[3] as their emission has been observed in a wide range of pressures and gas mixtures. High gain and fast decay is observed[4–6], with a timescale orders of magnitude shorter than the 67 nanosecond natural lifetime of the excited state[7].

In most experiments, the time dependence of the emission is studied by varying the time delay between an ultrashort "pump" pulse at 800 nm, followed by an ultrashort "probe" pulse which resonantly seeds the single photon transition[4–6,8–11]. This induces measurable changes in plasma radiation depending on the pump-probe delay, from which the temporal dynamics are inferred. However, this inevitably conflates unknown dynamics induced by the probe with the dynamics of interest. In addition, at each delay the reported measurement is integrated over the life time of the emission.

In this work, we use a streak camera to directly time-resolve fluorescence on the $B^2\Sigma_u^+(v=0) \leftarrow X^2\Sigma_g^+(v=0, v=1)$ transitions of $N_2^+$. In both transitions, the fluorescence emission profile exhibits 1) the characteristic delay with respect to the pump pulse and 2) decay time faster than the natural lifetime of the transition. Our observations agree with Dicke superfluorescence[12] previously observed with free electron lasers[13] and air lasing[6]. We emphasize that we observe this superfluorescence in air under ambient conditions, making it suitable for practical applications.

The contribution of collective emission in air lasing has been highlighted in pump probe studies of $N_2^+$ emission[14,15] in gas cells at low pressure. In the previous studies of a similar phenomena using streak camera some features of this collective emission such as delayed response has not been observed[16,17] or had been attributed to other sources[18].

We also show that the density profile of the generated plasma gives rise to spatially dependent emission temporal profile. That needs to be considered for proper engineering of a remote source for applications in backward emission.

## II. EXPERIMENTAL SETUP

In our experiment, bandwidth-limited ultrashort pulses of 50 fs duration, 1 mJ energy, at 1 kHz repetition rate centered at 800 nm are focused in air. In a tight focusing geometry (numerical aperture of 0.1), the peak intensity reaches $4 \times 10^{14}$ Watt/cm$^2$ generating a plasma that glows over one centimeter. The emission is self-seeded with the supercontinuum generated in the focused short pulse. Indeed, we observed the supercontinuum ring consisting of colorful rings covering the main beam in the far field.

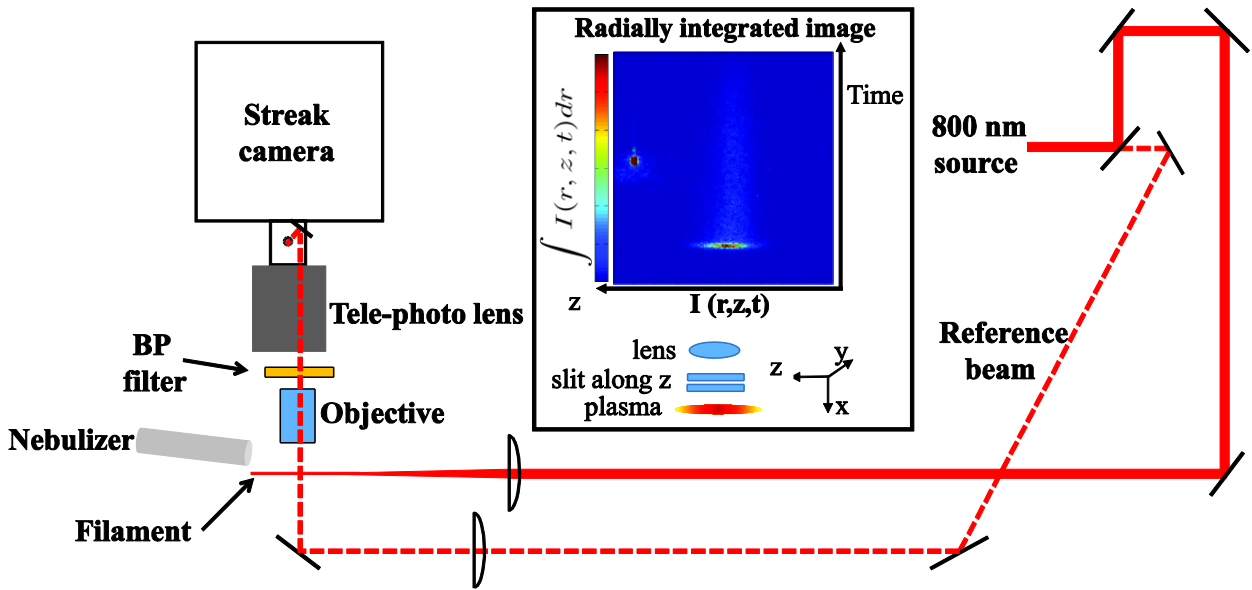

**FIG.1** The side emission of the plasma created by the focused main beam (thick red line) is imaged onto the streak camera. A weak beam split off from the main beam (dashed red line) provides a temporal reference spot on each frame of the streak camera. The streak camera slit is parallel to the main beam ($z$ direction) and is used at its minimum possible opening for best time resolution. A typical image taken with infrared filter is presented in the inset. The abscissa is the propagation axis, and the ordinate is time. The image is integrated over the transverse dimension as a function of $z$ (coordinate along the beam). The trace of the plasma radiation is recorded. The reference light (on the left) is used for timing reference and correction in spatial jitter.. BP represents the band-pass filter. A nebulizer is used to measure the Rayleigh scattering from the focused beam for a weak non-ionizing pulse.

The streak camera captures emission from the side of the filament, as opposed to integrating the emission along its length[4–6,8–10]. Thus, by making point by point measurements along the propagation direction, we access the "longitudinal emission profile" and can therefore monitor the effects of propagation .In order not to be limited by the electronic jitter of the streak camera with respect to the master clock of the laser, the images are synchronized by sending a weak beam through the slit of the streak camera[16] (see details in Appendix A). Rayleigh scattering from small droplets of distilled water created by a nebulizer is used to calibrate the position and timing of the emission (see details in Appendix B). We accumulate 1000 -2500 frames for each image to improve the signal-to-noise. For the sweep speed of 100ps used in all the data presented below, the spacing between pixels corresponds to 2.8767 ps.

We characterize the emission in three different spectral regions, selected by appropriate combination of filters: $391 \pm 1$ nm (corresponding to the $B^2\Sigma_u^+(v=0) \rightarrow X^2\Sigma_g^+(v=0)$ transition), $428 \pm 1$ nm (corresponding to the $B^2\Sigma_u^+(v=0) \rightarrow X^2\Sigma_g^+(v=1)$ transition), and $1\mu m > \lambda > 750$ nm that covers $X$ to $A$ transition. The time delay associated with each filter is accounted for. Note that the individual rotational transitions cannot be resolved.

The time dependence of the emission (integrated in space) of the plasma created by a focused 1 mJ pulse, is presented in the plots of Fig. 1. The time axis is calibrated by defining $t = 0$ as the arrival time of the Rayleigh scattering pulse. In addition, the Rayleigh scattering pulse serves as the instrument response function of our streak camera which we use to de-convolute the raw data and extract the true temporal profile of the fluorescence. Both raw and de-convoluted temporal emission profiles are plotted. The emission profiles of ionic lines do not follow the Voigt profile of the Rayleigh scattering profile, but are highly asymmetric, with a "fast" high gain peak, followed by a "slow" exponentially decaying tail[11,17] [Fig. 1].

We first turn to the slow decay dynamics. The infrared emission at $\lambda > 750$ nm (green curve) plotted in Fig. 1(c), has zero delay with respect to the Rayleigh scattering and decays exponentially with a time constant of 700 ps. The IR emission (green curve) in Fig 1(c) accompanies an identical decay behavior seen in both 391 and 428 nm emission profiles, suggesting the involvement of the $A$ state in the dynamics[19]: population exchange between $X$ and $A$ states affects gain at the $B$ to $X$ transition wavelength.

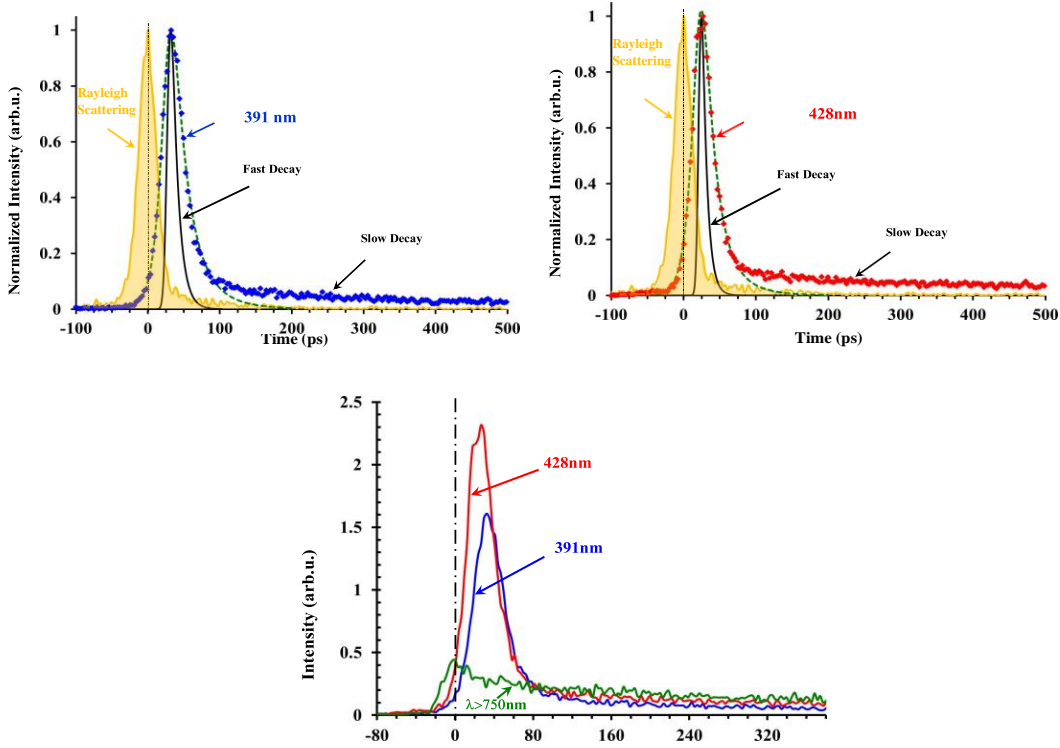

**FIG.2** Spatially integrated emission versus time. (a) and (b) show the normalized emission of the nitrogen cation at 391 nm (a) and at 428 nm (b), respectively. The yellow profile is the Rayleigh scattering from the exciting pulse at 800 nm. The raw data for both (a) and (b) is shown in blue and red respectively, the green dotted curve is the fit to data in proximity of the peak. The convoluted fast emission of 391 and 428 is calculated using the instrument response function of the Rayleigh scattering of the femtosecond pulse and is shown with black curve. (c) The 391 nm and 428 nm emission and broadband IR emission ($\lambda > 750$ nm) are plotted without normalization. There are two time constants for the ion emission known as "fast" and "slow" decay.[11,17]

Indeed, this exponential decay was not observed in the Rayleigh scattering measurement with low intensity pulse, in which $N_2^+$ is not generated.

We can estimate this slow decay assuming it arises from population decay due to electron- cation collisions as $1/\rho v \sigma$, where $\rho$ is the density of emitting dipoles, $v$ is the velocity of electrons and $\sigma$ is the collision cross section. $\rho = 0.01 \rho_{air}$ based on our average intensity of $10^{14}$ Watt/cm$^2$ which yields an ionization rate[20] of one percent. The average energy of released electrons is 0.1 eV with the total electron cross section[21] between $5 \times 10^{-16}$ cm$^2$. This yields a decay time of 700 picoseconds, in good agreement with the slow decay of the measured ion emission as well as the infrared emission in figure 1.

The fast peaked emission profiles exhibit characteristic delays of 33 ±0.3 ps at 391 nm [Fig.1(a)] and 25 ±0.3 ps for 428 nm [Fig.1(b)]. They also exhibit short temporal widths of 14 ±0.5 ps at 391 nm and 10 ±0.5 ps at 428 nm. The errors are the quadrature sum of the fitting and de-convolution errors.

Dicke superfluorescence predicts that these delay and decay times should be inversely pro- portional to the number of dipole emitters involved in the emission[12], which in turn is proportional to the time integral of the non-normalized emission profiles [Fig. 1(c)]. We find the ratio of the delay time for 391 nm emission to that of 428 nm to be 1.32 ± 0.02, while the ratio of temporal widths is 1.4 ± 0.09. Indeed, both these numbers are in close agreement with 1.4 ± 0.5, the ratio of number of emitters in the 428 nm to 391 nm emission profiles, obtained from their time integral. Note that this collective emission does not require an inverted sample[22].

The higher gain of the 428 nm fluorescence can be due to higher inversion of the $B^2\Sigma_u^+(v = 0) \rightarrow X^2\Sigma_g^+(v= 1))$ transition due to unequal occupation of the $X^2\Sigma^+$ vibrational states following strong-field ionization[23].

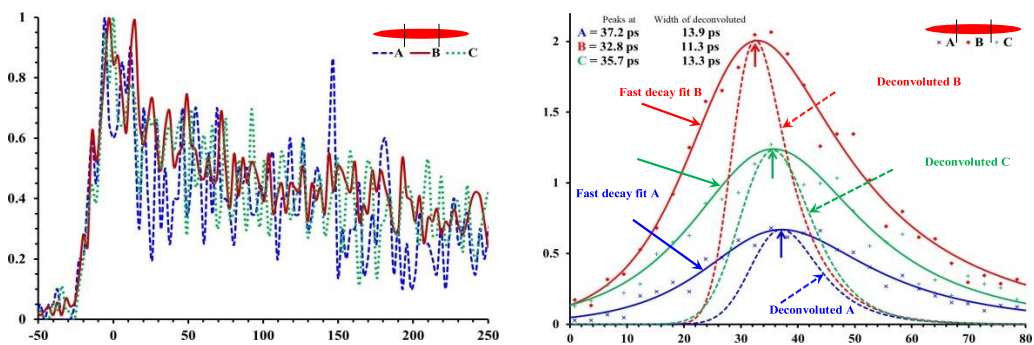

**FIG.3** Left : Integrated IR emission from selected spatial regions of the plasma labeled as "A", "B" and "C" as shown in the inset. Right: Comparing fast decay and built up time of the emission at 391 from the same selected regions.

In order to further investigate the collective contribution of emission, we analyze spatial properties of the measured fluorescence. The emission is divided in three equal slices along the propagation direction as shown in the inset of Fig 2. The emission is integrated in radial direction over each section and plotted as a function of streak camera sweep. Fig 2 shows the spectral emission with long pass filter (left) and the emission at 391 nm (right). The integrated signal in both IR and 391 is strongest in the central region of the plasma labeled as "B", followed by "C" and "A".

The normalized emission at IR shows that all regions have comparable time dependent profile and no observable delay between regions could be measured in our system. IR emissions from the selected regions exhibit their maximum at the Rayleigh peak. However a measurable difference is recorded for the peak emission at 391. The brightest region of the plasma has the least delay with respect to the Rayleigh peak, with a delay inversely proportional to the strength of the signal. The width of the emission increases as the total signal decreases. The 3 to 4.5 ps relative delay between emission originating from the center and sides of the plasma is much longer than the time it would take for light to travel between the sub-regions. We show that density of the emitters not only affect the strength of the signal but also the temporal profile of the emission. Altogether these observations suggest that it is important to include collective effects in modeling gain and propagation in air-lasing.

## III. CONCLUSION

Our unique capability of sub-picosecond timing resolution along with absolute timing measurement using the Rayleigh scattering enables us to report accurate measurement in timing and temporal profile of the emission from ($B^2\Sigma_u^+$, $v = 0$) → ($X^2\Sigma_g^+$, $v = 0$ & $v = 1$) in $N_2^+$. Our study suggests that the fast decay and high peak gain observed in air lasing is due to collective emission introduced by Dicke in 1954.[12]. The temporal behavior of the pulse as a function of density agrees with the predictions of superfluorescence[13, 14].

The presence of a long tail emission at infrared wavelength with no observable delay The presence of a long tail emission at infrared wavelength with no observable delay with respect to the initial pulse suggests the influence of *X* to *A* coupling in air lasing.

Our study suggests that collective emission and spatial distribution of the plasma needs to be considered for applications in remote source design using $N_2^+$ as a gain medium.

**FUNDING**

The Army Research office (W911NF-19-1-0272), U.S. Department of Energy (DOE) (DESC0011446) and the Air Force Research Laboratory (FA9451-15-1-0039).

**DISCLOSURE**

The authors declare that there are no conflicts of interest related to this article.

**ACKNOWLEDGMENTS**

We would like to acknowledge Andreas Velten, Paul Corkum. Michael Spanner, David Villeneuve, Mathew Britton and Andreas Schmitt-Sody for fruitful discussions on air lasing.

**Appendix A: Time resolution in streak camera imaging**

The emission the plasma is very weak, corresponding to a plasma density of the order of $10^{17}$ cm$^{-3}$. The total signal is further decreased by the fact that we attempt to measure a number of photons emitted per picosecond time gate with spatial resolution over the length of the plasma. One of the most difficult challenges to address is to collect as efficiently as possible all the light that is emitted radially. Various focusing geometries and slit sizes have been tested. Best results were obtained with a 10 cm focal distance lens, and by using an objective lens and an achromatic telephoto lens to image the transverse emission onto the slit of the streak camera. There are three elements that determine the time resolution of the system:

1. The accuracy of positioning the frames with respect to each other
2. The intrinsic resolution of the streak camera
3. The reproducibility of the event under observation

1. **A delta function reference point**

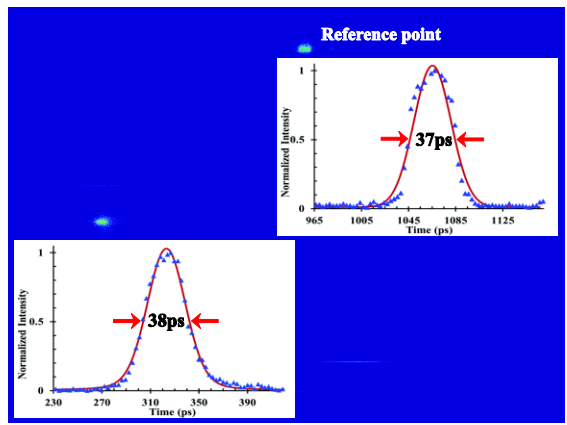

**FIG A:** Typical average of 1000 frames, recording the scattering of a diffuser. For ideal resolution, the reference (top) and the image have the same intensity. Note that they also have the same width.

There is a 60 nanosecond delay between trigger and sweep of the streak camera. Therefore one can not use the same laser pulse to trigger the camera for the same frame. The streak camera is triggered electronically through the master clock of the laser system, thereby inheriting jitters of the laser electronics. The accuracy of optical timing is transferred to each image by using a reference optical beam which is selected from the same pulse that creates the plasma and sent through a fixed path directly to the camera. The path of the reference beam is such that it always illuminates the same point of the photo-cathode in the time frame of the camera streak, providing a temporal reference for every streak camera image. Mechanical and electronic jitters between frames are corrected by using the timing of the reference pulse in the Matlab reconstruction code16. The reference beam was sent directly through a fiber connector on top of the streak camera, which resulted in an order of magnitude better resolution than when using the fiber provided by the streak camera manufacturer.

2. **Imaging the laser focal spot; time and space reference**

The response to a delta function light pulse, obtained by scattering the filament off a diffuser, is extremely sensitive to the slit opening. In fact, it is only with the slit opened to its minimum (i.e. nearly closed) that we achieve the resolution quoted by the manufacturer. The signal is then so weak that it is reduced to a few scattered dots. In order to achieve the best resolution,it is necessary to accumulate and average between 1000 and 2500 frames. In doing so, the frames have to be synchronized as discussed in the previous paragraph. The intensity of the synchronization pulse should also be such that only 10 to 20 pixels are irradiated per frame.

The algorithm takes the average of all the synchronization dots, excluding those that are outside of an area equal to 4 times the mean square deviation. This average is taken as reference for all frames. Figure A shows an image of a scatterer (1000 frames are averaged). Both reference point and image have the same width of 38 ps.

This picture indicates that we can define the centroid and the rise time with a precision of a few ps.

# Appendix B: Absolute timing of the nitrogen cation emission

Rayleigh scattering at 800 nm from air molecules is too weak to be observed (and time resolved) with the streak camera. Since we have achieved accurate timing of each frame with the reference pulse as demonstrated in Fig B we can perform independent measurements to determine the arrival time of the 800 nm pulses.

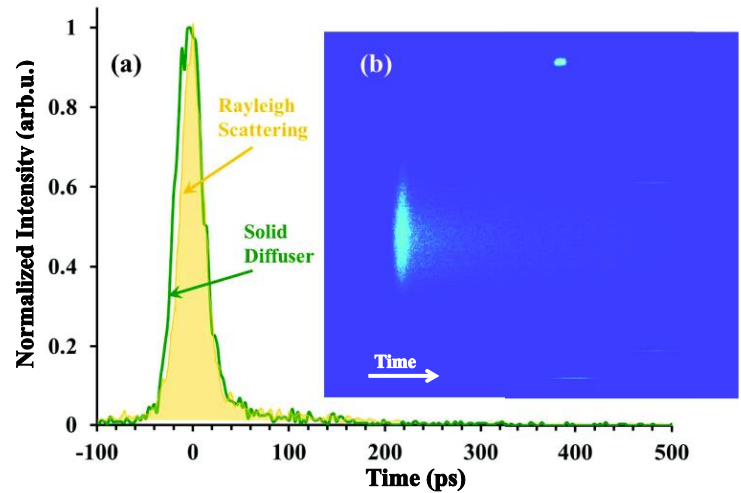

**FIG B**. (a) In yellow: the reference pulse obtained by sending the laser beam through a dispersed mist. Green: reference pulse obtained with a solid diffuser surface. The dashed line indicating the center of gravity of the yellow reference will be used as the time of arrival of the fs pulse. (b) typical streak camera image frame.

The problem is to ascertain with precision that exactly the same spot is being observed in both experiments. Using a diffuser plate - such as was done in the experiment to determine the instrumental resolution - puts one at the mercy of an error in positioning. The solution that we chose isuse Rayleigh scattering enhanced by aerosols.

The aerosols are blown through the camera field of view along the path of the focused beam. The challenge here is to create droplets that enhance scattering without creating plasma and/or producing and optical resonances[24] in the droplet.

The solution chosen is to produce droplets of the order of 1 $\square$m diameter with a nebulizer (mist generator based on Bernoulli principle). These droplets were sufficiently small as not to create any visible plasma or local illumination. The reference point accurately determined by recording Rayleigh scatteringof the lowest density mist is shown in Fig B.